\documentclass[fleqn,12pt,twoside,A4]{article}
\usepackage[headings]{espcrc1}

\readRCS
$Id: espcrc1.tex,v 1.2 2004/02/24 11:22:11 spepping Exp $
\usepackage{graphicx}
\usepackage[figuresright]{rotating}

\newcommand{\AmS}{{\protect\the\textfont2
  A\kern-.1667em\lower.5ex\hbox{M}\kern-.125emS}}

\newcommand{\jpsi}{J/\psi}

\title{Probing the properties of dense partonic matter at RHIC}

\author{Y. Akiba\address[RIKEN]{RIKEN, The Institute of Physical and Chemical
Research, Wako, Saitama 351-0198, Japan},
for the PHENIX Collaboration\footnote{for the full list of PHENIX authors 
and acknowledgements, see Appendix ``Collaboration'' of this volume.
}}

\runtitle{Probing the properties of dense partonic matter at RHIC}
\runauthor{Y. Akiba for the PHENIX Collaboration}

\begin{document}
\maketitle

\begin{abstract}
The experimental highlights from PHENIX are summarized from the point of
view of probing the properties of dense partonic matter produced at RHIC.
\end{abstract}

\section{Introduction}
The first three years of RHIC results have been summarized by the four RHIC
experiments in their White Papers\cite{PHENIX_WP,BRAHMS_WP,PHOBOS_WP,STAR_WP},
which were published during this conference.
It is appropriate to start the summary and the focus of the results of PHENIX
in this Quark Matter 2005 conference from the conclusions of the white paper.
In the last paragraph of the PHENIX white paper we wrote\cite{PHENIX_WP}:
\begin{quotation}
{\em
In conclusion, there is compelling experimental evidence that
heavy-ion collisions at RHIC produce a state of matter characterized
by very high energy densities,
...\\
... additional incisive
experimental measurements combined with continued refinement of the
theoretical description is needed to achieve a complete understanding
of the state of matter created at RHIC.
}
\end{quotation}
The questions we want to address in this summary are:
\begin{itemize}
\item {\em What additional incisive experimental measurement do we now have?}
\item {\em What additional information on the properties of the matter can
be derived from the new data?}
\end{itemize}

\section{Highlights of new PHENIX data}
\begin{figure}[ht]
\vspace*{-5mm}
\begin{center}
\includegraphics[width=0.7\linewidth]{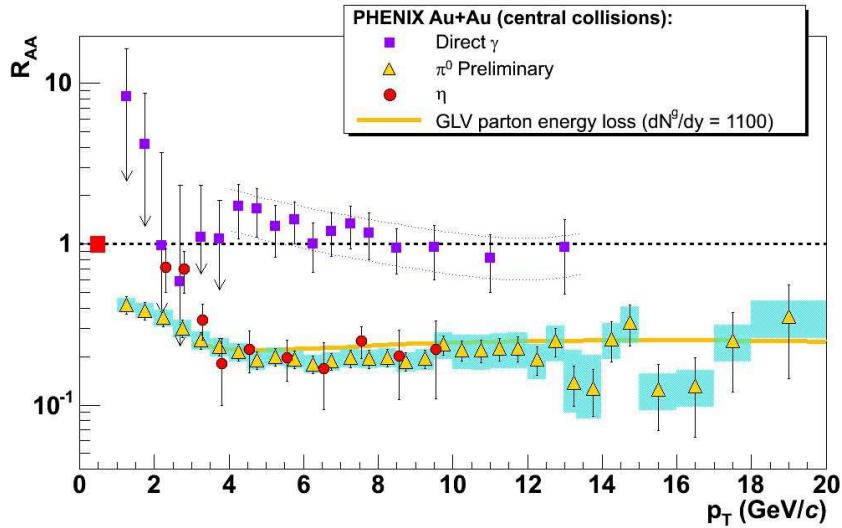}
\end{center}
\vspace*{-15mm}
\caption{\label{fig:RAA_pi0} Nuclear modification factor, $R_{AA}$ of $\pi^0$
(triangles), $\eta$ (circles), and direct photon (squares).}
\end{figure}

We have observed that the matter is very opaque and dense.
It is so dense that even a 20 GeV/$c$ pion is stopped.
In figure~\ref{fig:RAA_pi0} we show our preliminary data for the nuclear
modification factor, $R_{AA}$, of $\pi^0$ in central
Au+Au collisions in the $p_T$ range up to 20 GeV/$c$\cite{MAYA}.
The suppression
is very strong, and it is flat at $R_{AA} \simeq 0.2$ up to 20 GeV/$c$. 
There is no hint that it returns to unity.
The figure also shows that the suppression of $\pi^0$'s and $\eta$'s is very
similar, which supports the conclusion that the 
suppression occurs at the parton level, not the hadron level.
This strong suppression of mesons is in stark contrast to the behavior of
direct photons\cite{PPG042},
also shown in the figure. The direct photons follow
binary scaling (i.e. $R_{AA} \simeq 1$). This is strong evidence that the
suppression is not
an inital state effect, but a final state effect caused by the high density
medium created in the collision.
The curve in the plot shows a theoretical prediction\cite{GV}
using the GLV parton energy loss model. The model
assumes an inital parton density $dN/dy=1100$, which corresponds to an energy
density of approximately 15 GeV/fm$^3$. The data show
that the suppression is somewhat stronger than the prediction, suggesting that
the matter density may be even higher than these estimates.

\begin{figure}[ht]
\begin{minipage}[t]{75mm}
\vspace*{-5mm}
\includegraphics[width=\linewidth]{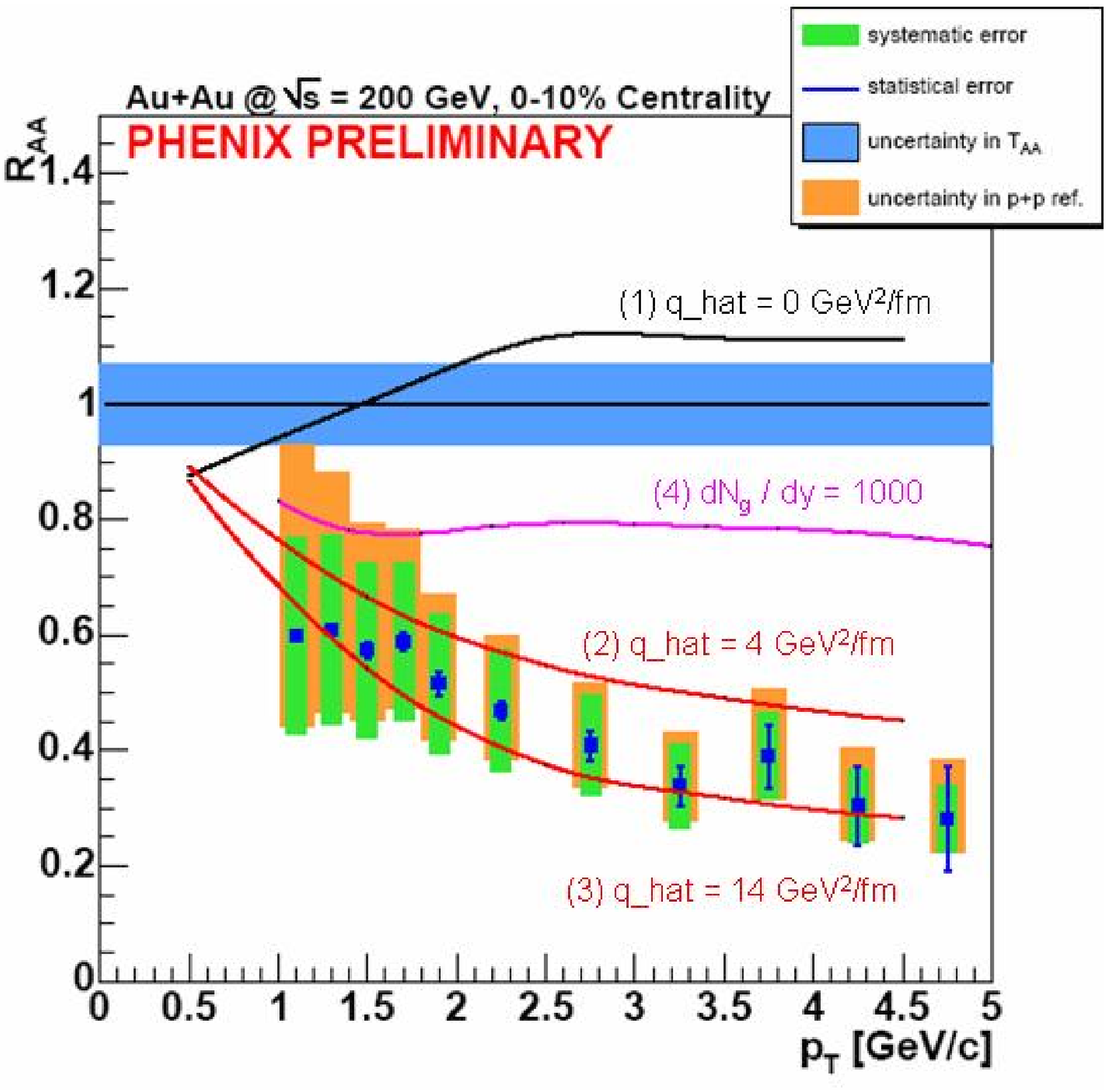}
\vspace*{-14mm}
\caption{\label{fig:RAA_e} The nuclear modification factors $R_{AA }$
of single electrons from heavy quark decay. The theory
predictions ((1)-(3) from \cite{Armestro} and (4) from \cite{MDjordjevic})
are also shown.}
\end{minipage}
\hspace{\fill}
\begin{minipage}[t]{75mm}
\vspace*{-5mm}
\includegraphics[width=\linewidth]{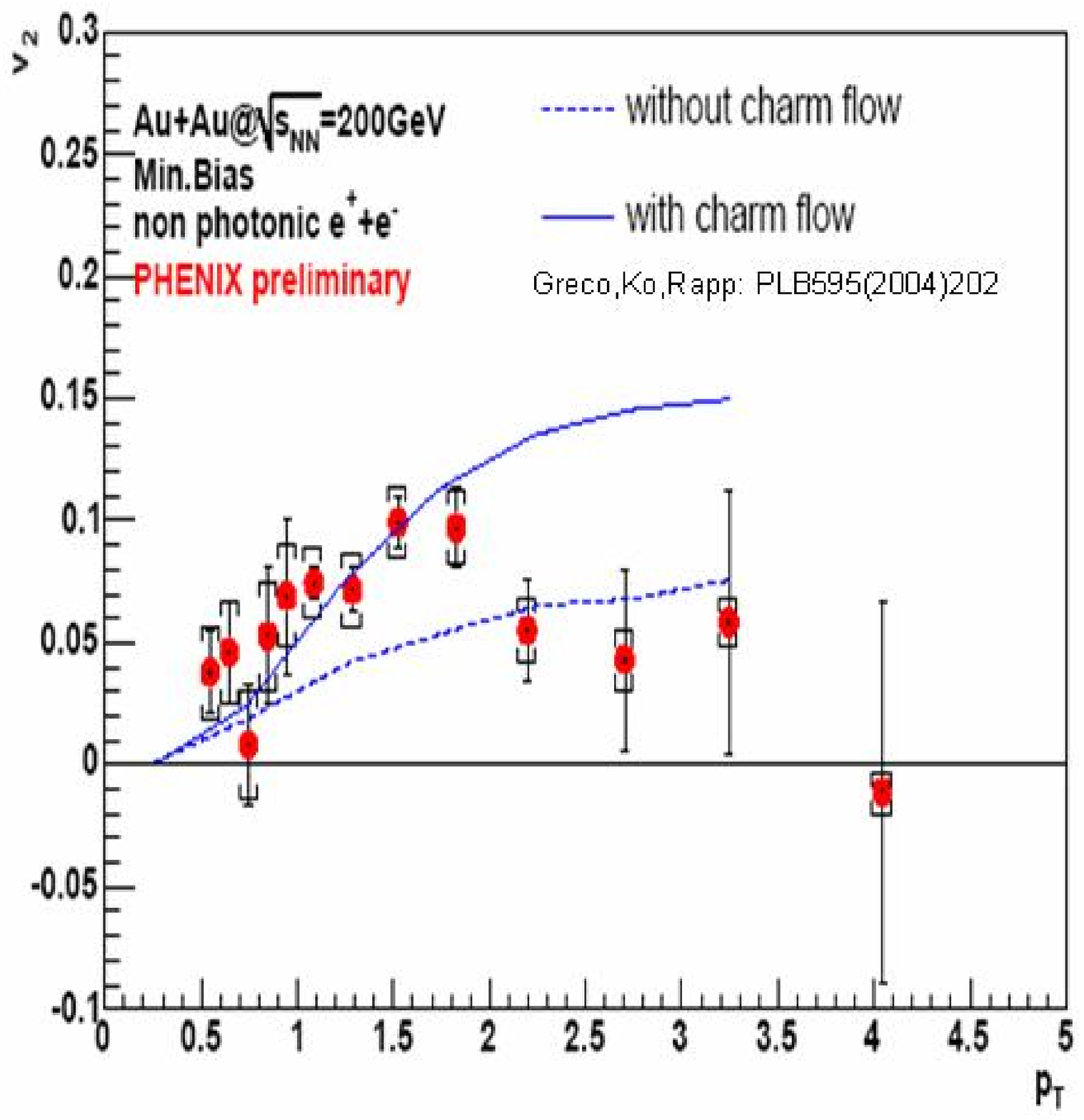}
\vspace{-17mm}
\caption{\label{fig:ev2} The elliptic flow strength $v_2$
 of single electrons from heavy quark decay. The curves on the figure are
charm coalescence model predictions\cite{Greco} with (solid) and
without (dashed) charm quark flow.}
\end{minipage}
\end{figure}

We have observed that the matter is so dense that even heavy quarks are
stopped. The data shown in figure~\ref{fig:RAA_e} are the nuclear modification
factors, $R_{AA}$, of single electrons from heavy flavor decay.
The data show that heavy quarks, mainly charm in this $p_T$ region, suffer
substantial energy loss in the matter, which results in strongly
suppressed single electron spectra. The suppression is very strong, almost as
strong as that of light flavor mesons ($\pi^0$ and $\eta$).

These data provide strong constraints on and challenges to the energy loss
models. The curves in the figure are theoretical
predictions\cite{Armestro,MDjordjevic}.
The theory curves of \cite{Armestro} are for charm quarks only, while
that of \cite{MDjordjevic} includes the effect of beauty, which is predicted
to have smaller energy loss in the medium. If the effect of beauty is removed,
the predictions of the two approaches are quite similar.
The strong suppression shown in the data requires a large 
transport coefficient (as large as {\it \^{q}} = 14 GeV$^2$/fm),
or correspondingly a very high initial parton density
(as high as $dN/dy = 3500$). Such a high parton density may be consistent
with the strong suppression observed in $\pi^0$, but it is a challenge to
the theories since it is not compatible with the observed final state
particle multiplicity. In addition,
the data may require a strong energy loss of beauty in the medium. This leads
to a recent suggestion that the main energy loss mechanism is not gluon
radiation, but elastic scattering\cite{Rapp,Gyulassy}.

This matter is so strongly coupled that even heavy quarks flow.
Figure~\ref{fig:ev2} shows our preliminary data of the elliptic flow
strength, $v_2$, of single electrons from heavy quark decay\cite{Sergey}.
The data clearly demonstrates that the $v_2$ of single electrons is non-zero,
and that therefore the parent $D$ meson have non-zero elliptic flow.
This confirms our earlier results\cite{PPG040}.
In order to get some idea of how strong the $D$ meson $v_2$ is, 
we compare the data with the calculated single electron $v_2$ from the
$D$ meson decay. Here we use a simple {\it ansatz} that the $v_2(p_T)$ of
the $D$ mesons has the same shape as that of pion, but its strength is
scaled. The comparison shows that our data can be reproduced rather well if the
$D$ meson $v_2$ is approximately 60\% that of pion. Thus the charmed mesons
flow, but not as strongly as light mesons. 
We also see that the flow strength drops in higher $p_T$, above 2.5 GeV/$c$.
This is the $p_T$ region where the contribution from
beauty decays becomes significant. Is the drop due to the $B$ meson
contribution?
This is an open question which has to be addressed in future.
In the figure, we compare the theoretical predictions of \cite{Greco} with
our data. The data at low $p_T$ favors the models that include flow for the
charm quarks. In turn, charm quark flow would indicate high parton density
and strong coupling in the matter. The strength of the flow can be reproduced
either by resonance-like states\cite{Rapp} or a large cross section of
$c$-quarks ($\sigma \simeq 10$ mb)\cite{Zhang} in the dense medium. If these
models are correct, the charm flow data can be taken as evidence for strongly
coupled QGP. It is not a weakly coupled gas.

\begin{figure}[h]
\begin{minipage}[t]{65mm}
\vspace*{-5mm}
\includegraphics[width=\linewidth]{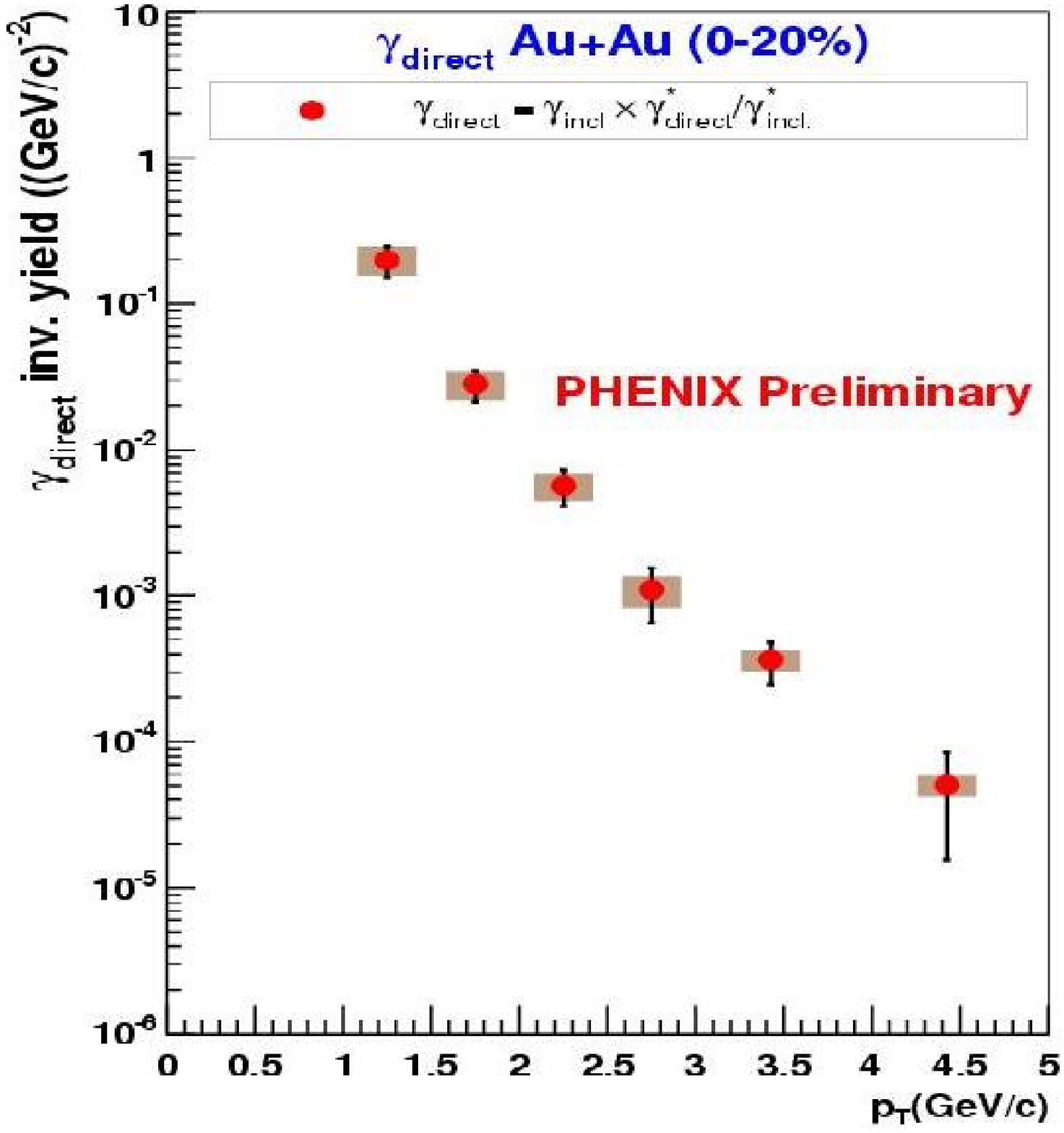}
\vspace*{-14mm}
\caption{\label{fig:photon} The transverse momentum spectrum of intermediate
$p_T$ direct photons from central Au+Au collisions at $\sqrt{s_{NN}}=200$ GeV.}
\end{minipage}
\hspace{\fill}
\begin{minipage}[t]{85mm}
\vspace{-2mm}
\includegraphics[width=\linewidth]{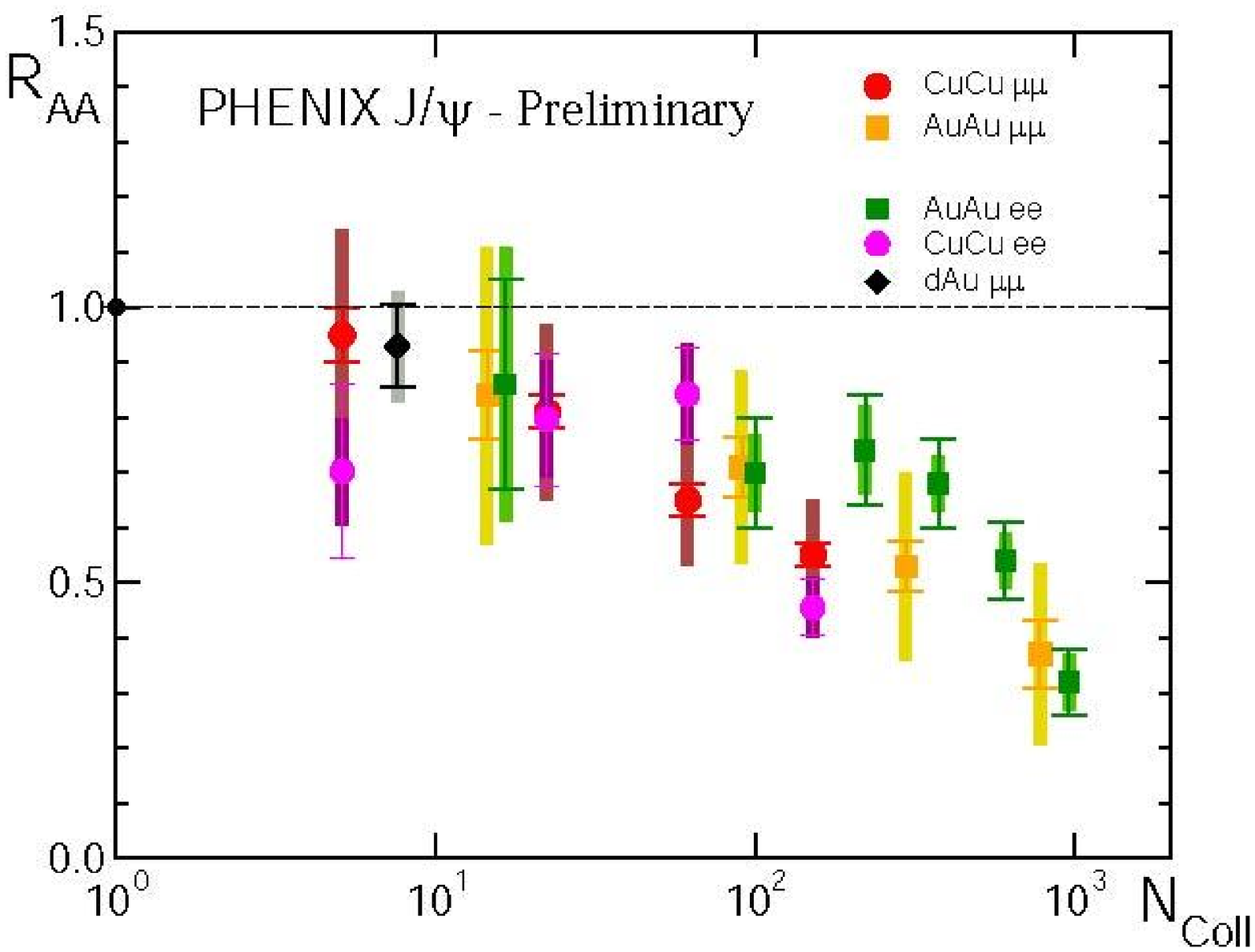}
\vspace*{-16mm}
\caption{\label{fig:RAA_jpsi} The nuclear modification factors
$R_{AA}$ of $\jpsi$ in $d+$Au, Cu+Cu, and Au+Au collisions
at $\sqrt{s_{NN}}=200$ GeV.}
\end{minipage}
\end{figure}

We now have the first promising result of direct photon measurements at
intermediate $p_T$ ($1 < p_T < 5$ GeV/$c$) from the analysis of low mass,
high $p_T$  electron pairs\cite{Bathe}. In figure~\ref{fig:photon}, we show
our preliminary direct photon results.
The measurement is done through low mass electron pair analysis, but it is
converted to the real photon invariant yield assuming
$\gamma_{direct}/\gamma_{incl.} = \gamma^{*}_{direct}/\gamma^{*}_{incl.}$.
Are these thermal photons? The rate is above a pQCD calculation\cite{WV}
scaled by $T_{AB}$. The pQCD calculation agrees with our direct photon
measurements in $p+p$ very well in $p_T>5$ GeV/$c$.
However, we need reference measurements in $p+p$ and
$d+$Au to draw definite conclusions. The same analysis method can be used in
$p+p$ to test the validity of pQCD in this low $p_T$ region, and in $d+$Au
to examine cold nuclear matter effects such as Cronin effect. If the yield is
due to thermal radiation, the data can provide the first direct measurement
of the initial temperature of the matter through comparison between the data
and theoretical calculations. Comparison with a model
prediction\cite{PHOTON_MODEL} yields the value of 500 to 600 MeV for
$T_{max}$, where $T_{max}$ is the maximum temperature at the center of
the fireball. Averaging over the entire volume leads to smaller values
of 300 to 400 MeV. These values are, of course, only meaningful if and only
if these photons are indeed thermal photons, but nevertheless these results
are quite intriguing.

We now have very comprehensive data of $\jpsi$ production at RHIC energies,
from $p+p$ to $d+$Au to Cu+Cu to Au+Au; both in central rapidity ($|y|<0.35$)
in the $e^+e^-$ channel and in forward rapidity ($1.2<|y|<2.4$) in the
$\mu^+\mu^-$ channel\cite{Hugo}. The nuclear modification factors, $R_{AA}$, of $\jpsi$
at RHIC energies are presented in figure~\ref{fig:RAA_jpsi}.
The data show that, in the most central collision, $\jpsi$'s 
are clearly suppressed beyond the cold nuclear effect. The preliminary data
are consistent with predictions of the models with the suppression
plus re-generation through charm recombination/coalescense\cite{JPSI_RECO}.
Do the data really show suppression plus re-generation? 
One way to test the re-generation model is to measure the $v_2$ of $\jpsi$.

\begin{figure}[h]
\vspace*{-5mm}
\begin{center}
\includegraphics[width=0.7\linewidth]{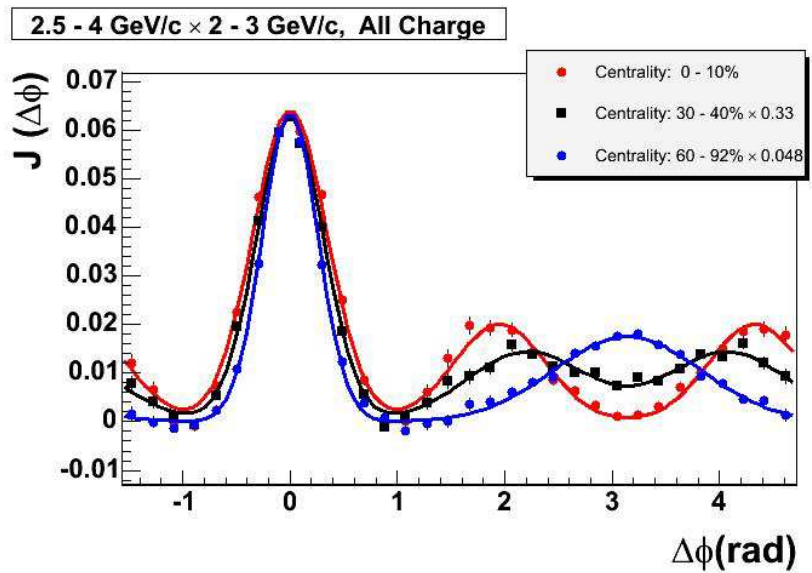}
\end{center}
\vspace*{-15mm}
\caption{\label{fig:jet_mod} Jet modification in Au+Au collisions with
different centralities.}
\end{figure}

We have observed that the shape of the jets is modified by the matter.
Figure~\ref{fig:jet_mod} shows our preliminary data of jet shape
analysis\cite{Grau}. The data show that the shape of the away-side jets changes
with the centrality. The reason for the modification of the jet shape 
is not well understood. It could be exotic effect like Mach
cone\cite{Mach}, or Cerenkov-like radiation\cite{Wang}, or something that
is more conventional. If this is due to Mach cone or Cerenkov-like radiation,
the jet shape measurements will open up new possibilities to measure
the properties of the matter. Can the properties such as the velocity of sound
or color di-electric constant be measured from the shapes? The data show great
potential for the di-jet tomography as a powerful tool to probe the properties
of the matter.

\begin{figure}[ht]
\vspace*{-5mm}
\begin{center}
\includegraphics[width=\linewidth]{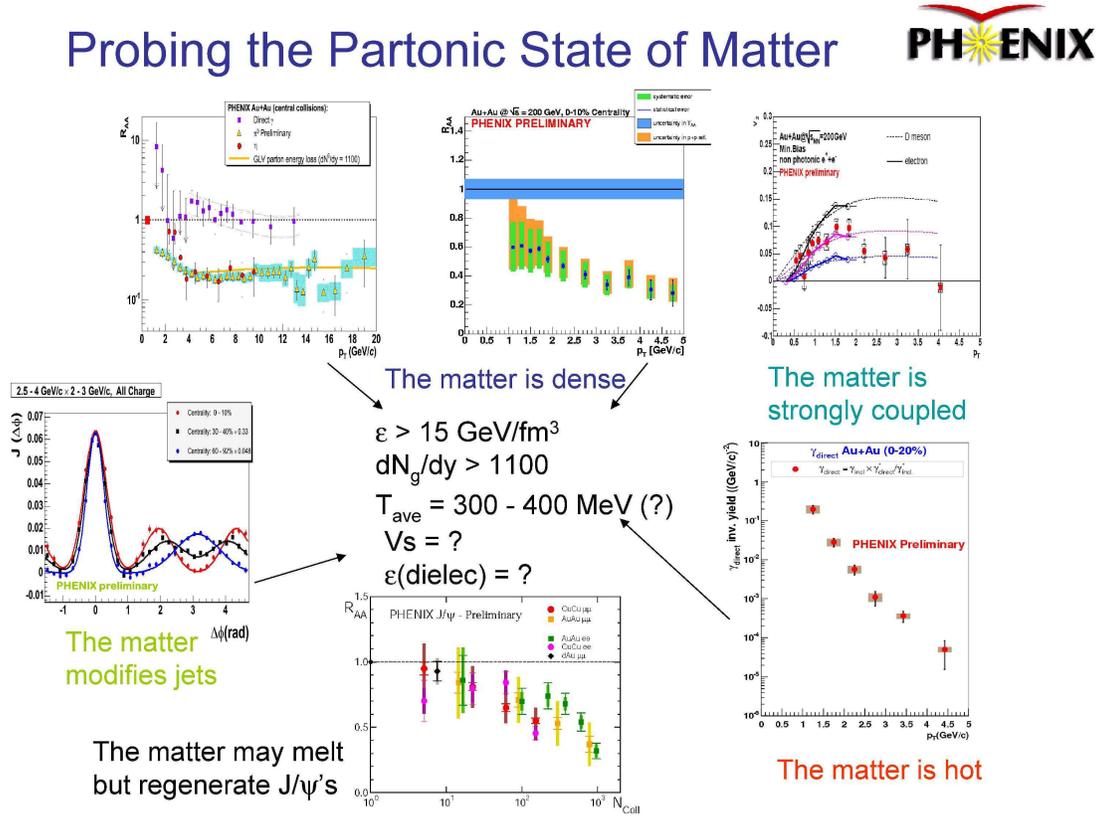}
\end{center}
\vspace*{-15mm}
\caption{\label{fig:circle} Probing the partonic state of matter created at
RHIC from many observables.}
\end{figure}

\section{Summary and Conclusions}
We are now moving from the initial discovery phase of dense partonic matter
to the next phase of probing the properties of the matter created at RHIC.
The penetrating probes such as high $p_T$ particles, heavy quarks, direct
photons, $\jpsi$'s, and jet tomography are very powerful tools to
study the properties of the medium, as summarized symbolically in
figure~\ref{fig:circle}. The new PHENIX preliminary data show that
the matter is so dense and opaque that even 20 GeV/$c$ pions and heavy quarks
are stopped; it is so strongly coupled that even charm quarks flow; it might be
so hot that it copiously produce thermal radiation; it may melt but regenerate
$\jpsi$; and it modifies jets. Close cooperation between theory and experiment
is needed to quantitatively determine the properties of the medium.
We look forward to working with the theory community to extract the
properties of the matter.

\end{document}